\documentclass[12pt]{article}
\begin{document}

\begin{center}
{\LARGE \bf The language of Einstein spoken \\ by optical instruments}

\vspace{7mm}

Sibel Ba{\c s}kal \footnote{electronic
address:baskal@newton.physics.metu.edu.tr} \\
Department of Physics, Middle East Technical University,\\
06531 Ankara, Turkey
\vspace{5mm}

Y. S. Kim\footnote{electronic address: yskim@physics.umd.edu}\\
Department of Physics, University of Maryland,\\
College Park, Maryland 20742, U.S.A.\\

\end{center}

\begin{abstract}
Einstein had to learn the mathematics of Lorentz transformations
in order to complete his covariant formulation of Maxwell's equations.
The mathematics of Lorentz transformations, called the Lorentz group,
continues playing its important role in optical sciences.  It is the
basic mathematical language for coherent and squeezed states.
It is noted that the six-parameter Lorentz group can be represented
by two-by-two matrices.  Since the beam transfer matrices in ray
optics is largely based on two-by-two matrices or $ABCD$ matrices,
the Lorentz group is bound to be the basic language for ray optics,
including polarization optics, interferometers, lens optics,
multilayer optics, and the Poincar\'e sphere.  Because the group of
Lorentz transformations and ray optics are based on the same two-by-two
matrix formalism,  ray optics can perform mathematical operations
which correspond to transformations in special relativity.  It is
shown, in particular, that one-lens optics provides a mathematical
basis for  unifying the internal space-time symmetries of massive
and massless particles in the Lorentz-covariant world.
\end{abstract}

\section{Introduction}\label{intro}
Before formulating his special relativity in 1905, Einstein studied
Maxwell's equations and concluded that classical electromagnetic
theory is consistent with the Lorentz-covariant space and time, instead of
the Galilean world on which Newton's mechanics is based~\cite{sommer52}.
When he wrote down Newton's $\vec{f} = m \vec{a}$ with the Lorentz force
for $\vec{f},$ which is covariant under Lorentz transformations,
Einstein to had to make $m \vec{a}$ Lorentz-covariant.

Indeed, Einstein introduced to physics the mathematics of Lorentz
transformations formulated earlier by Henri Poincar\'e, which is known
today as the Lorentz group.  In addition, as is well known, Einstein
established the relation between the photon energy and the frequency of
the associated light wave.  In so doing, Einstein
showed that photons are massless particles in the Lorentz-covariant world.

Newton had to invent a new mathematics now called calculus to formulate
his physical laws.  Newton's mathematical influence is not restricted to
physics.  Calculus is now an important scientific language even economics,
biology, and behavioral science.  Likewise, the Lorentz group, which
Einstein used for his relativity, serves as the underlying
scientific language for many different fields of physics, including
quantum field theory, the phase-space picture of classical and quantum
mechanics, and theories of superconductors.

Recently Einstein's influence is becoming more prominent in optical
sciences.  It is by now well established that
coherent and squeezed states are harmonic-oscillator representations
of the Lorentz group~\cite{knp91}.  More recently, the Lorentz group
has been shown to be the underlying scientific language for classical
ray optics.  The group of Lorentz transformations consists of
four-by-four matrices applicable to the four-dimensional Minkowskian
space-time.  However, it is mathematically possible to represent the
same Lorentz group using much simpler two-by-two matrices.

Since classical ray optics is mostly based on two-by-two matrices, it
is essentially the physics of the Lorentz group, as special relativity
is.  It is remarkable that this physics of two-by-two matrices
embraces two completely separate branches of physics.  It is
straight-forward to rewrite the Jones-matrix formalism in terms of
the Lorentz group~\cite{hkn97}.  Since it is possible to construct
mathematically the four-by-four representation from the two-by-two
representation, the four-parameter Stokes parameters form a Minkowskian
four-vector, on which Einstein's special relativity is based~\cite{hkn99}.

Para-axial lens optics is also based on two-by-two matrices, so are the
optical rays in laser cavities.  Thus, both can be regarded as the
physics of the Lorentz group.  This group allows us to derive some powerful
results in lens optics and laser cavities~\cite{bk01,bk02,bk03}. Multilayer
optics involving reflections and transmissions is also the physics of
two-by-two matrices~\cite{azzam77,monzon00}.  Here also, the Lorentz
group can play a fundamental role~\cite{gk01,gk03}.  These latest
developments in ray optics have been summarized in a recent review
paper~\cite{bgkn04}.

The simplest matrices, next to one-by-one, are two-by-two matrices.
What more is there to learn?  Yes, they are simple to deal with
if there are two or three two-by-two matrices.  If there are more,
calculations become tedious and uncontrollable.  We need group theory
to deal with systematically those complicated matrix multiplications.
In addition, those matrices speak Einstein's language for special
relativity.  By arranging optical instruments, we can perform
the mathematics corresponding to Lorentz transformations.  Compared
with those transformations performed high-energy
laboratories, optics experiments are very inexpensive.

In Sec.~\ref{polari}, we illustrate how polarization optics naturally
accommodates the language of the Lorentz group.  In Sec.~\ref{olens},
we illustrate how one-lens optics, with three two-by-two matrices,
can perform the calculation of group contractions which corresponds
to unification of internal space-time symmetries of massive and
massless particles.

\section{Polarization Optics}\label{polari}
Let us consider two optical beams propagating along the $z$ axis.
We are then led to the column vector:
\begin{equation}\label{jones}
\pmatrix{A_{1}~\exp{\left(-i\left(kz -\omega t + \phi_{1}\right)\right) }
\cr A_{2}~\exp{\left(-i\left(kz -\omega t + \phi_{2}\right)\right) } } .
\end{equation}
We can then achieve a phase shift between the beams by applying the
two-by-two matrix:
\begin{equation}\label{rot11}
\pmatrix{e^{i\phi/2} & 0 \cr 0 & e^{-i\phi/2}}.
\end{equation}
If we are interested in mixing up the two beams, we can apply
\begin{equation}\label{rot22}
\pmatrix{\cos(\theta/2)  & -\sin(\theta/2)  \cr
\sin(\theta/2)  & \cos(\theta/2)}
\end{equation}
to the column vector.

If the amplitudes become changed by either by attenuation or
reflection, we can use the matrix
\begin{equation}\label{boost11}
\pmatrix{e^{\eta/2}  & 0  \cr 0 & e^{-\eta/2}}
\end{equation}
for the change.  In this paper, we are dealing only with the
relative amplitudes, or the ratio of the amplitudes.

Repeated applications of these matrices lead to the form
\begin{equation}\label{alpha}
\pmatrix{\alpha & \beta \cr \gamma  & \delta} ,
\end{equation}
where the elements are in general complex numbers.  The
determinant of this matrix is one.  Thus, the matrix can
have six independent parameters.

Indeed, this matrix is the most general form of the matrices
in the $SL(2,C)$ group, which is known to be the universal
covering group for the six-parameter Lorentz group.  This means
that, to each two-by-two matrix of $SL(2,C)$, there corresponds
one four-by-four matrix of the group of Lorentz transformations
applicable to the four-dimensional Minkowski space.\cite{knp86}
It is possible to construct explicitly the four-by-four
Lorentz transformation matrix from the parameters $\alpha, \beta,
\gamma,$ and $\delta$.  This expression is available in the
literature,\cite{knp86} and we consider here only special cases.

We can translate the above two-by-two matrices into their four-by-four
counterparts applicable to the four-dimensional Minkowskian space-time
$(ct, z, x, y)$.  The phase shift matrix of Eq.(\ref{rot11}) corresponds
to
\begin{equation}\label{rot411}
\pmatrix{1 & 0 & 0 & 0 \cr 0 & 1 & 0 & 0 \cr
0 & 0 & \cos\phi & -\sin\phi \cr 0 & 0 & \sin\phi & \cos\phi} ,
\end{equation}
and the rotation matrix of Eq.(\ref{rot22}) to
\begin{equation}\label{rot422}
\pmatrix{1 & 0 & 0 & 0 \cr 0 & \cos\theta & -\sin\theta & 0 \cr
0 & \sin\theta & \cos\theta & 0 \cr 0 & 0 & 0 & 1} .
\end{equation}
Repeated applications of these two matrices with different angle
parameters will lead to the most general form of the three-dimensional
rotation matrix applicable to the three-dimensional space of
$(z, x, y)$~\cite{gold80}.

As for the attenuation matrix of Eq.(\ref{boost11}), the corresponding
four-by-four matrix is
\begin{equation}\label{boost411}
\pmatrix{\cosh\eta & \sinh\eta & 0 & 0 \cr \sinh\eta & \cosh\eta & 0 & 0 \cr
0 & 0 & 1 & 0 \cr 0 & 0 & 0 & 1},
\end{equation}
which performs a Lorentz boost along the $z$ direction.  Repeated
applications of the above three four-by-four matrices lead to the most
general form for the Lorentz-transformation matrix.

If the Jones matrix contains all the parameters for
the polarized light beam, why do we need the mathematics in
the four-dimensional space?  The answer to this question is
well known.  In addition to the basic parameter given by the
Jones vector, the Stokes parameters give the degree of
coherence between the two rays.

Let us write Eq.(\ref{jones}) as  a Jones spinor of the form
\begin{equation}
 \pmatrix{\psi_{1}(z, t) \cr \psi_{w}(z, t) } ,
\end{equation}
Then the Stokes vector consists of
\begin{eqnarray}
&{}&  S_{0} = <\psi_{1}^{*}\psi_{1}>  + <\psi_{2}^{*}\psi_{2}> , \quad
 S_{1} = <\psi_{1}^{*}\psi_{1}>  - <\psi_{2}^{*}\psi_{2}> ,
  \nonumber \\[2ex]
&{}&  S_{2} = <\psi_{1}^{*}\psi_{2}> + <\psi_{2}^{*}\psi_{1}> , \quad
  S_{3} = -i\left(\psi_{1}^{*}\psi_{2}> - <\psi_{2}^{*}\psi_{1}>\right) .
\end{eqnarray}
The four-component vector $(S_{0}, S_{1}, S_{2}, S_{3})$
 transforms
like the four-vector $(t, z, x, y)$ under Lorentz transformations.
The Mueller matrix is therefore like the Lorentz-transformation
matrix.

Why do we need this Stokes four-vector, in addition to the Jones
spinor?  The Stokes parameters can deal with coherence between the
two independent beams.  As in the case of special relativity,
let us consider the quantity
\begin{equation}
M^{2} = S_{0}^{2}- S_{1}^{2} - S_{2}^{2} - S_{3}^{2} .
\end{equation}
Then $M$ is like the mass of the particle while the Stokes four-vector
is like the four-momentum.

If $M = 0$,  the two-beams are in a purely state. As $M$ increases,
the system becomes mixed, and the entropy increases. If it reaches the
value of $S_{0}$, the system becomes completely random.  It is
gratifying to note that this mechanism can be formulated in terms of
the four-momentum in particle physics.\cite{hkn99}

\section{One-lens System}\label{olens}
In analyzing optical rays in para-axial lens optics, we start with
the lens matrix and the translation matrix written as
\begin{equation}\label{lens}
 L = \pmatrix{1 & 0 \cr -1/f & 1} , \qquad
 T = \pmatrix{1 & z \cr -0 & 1} ,
\end{equation}
respectively.
Then the one-lens system consists of
\begin{equation}
\pmatrix{1 & z_{2} \cr 0 & 1}
\pmatrix{1 & 0 \cr -1/f & 1}
\pmatrix{1 & z_{1} \cr 0 & 1} =
\pmatrix{1 - z_{2}/f   & z_{1} + z_{2} - z_{1}z_{2}/f   \cr
-1/f & 1 - z_{1}/f } .
\end{equation}
If we assert that the upper-right element be zero,  then
\begin{equation}
{1 \over z_{1}} + {1 \over z_{2}} = {1 \over f} ,
\end{equation}
and the image is focussed, where $z_{1}$ and $z_{2}$ are the distance
between the lens and object and between the lens and image respectively.
They are in general different, but we shall assume for simplicity that
they are the same: $z_{1} = z_{2} = z$.  We are doing this because this
simplicity does not
destroy the main point of our discussion, and because the case with
two different values has been dealt with in the literature~\cite{gk03}.
Under this assumption, we are left with
\begin{equation}
\pmatrix{1 - z/f  &  2z - z^{2}/f  \cr  -1/f  & z/f -1 } ,
\end{equation}
which can be renormalized to
\begin{equation}\label{core}
C = \pmatrix{ x - 1  & x -2  \cr x  & x - 1} ,
\end{equation}
with $x = z/f$.

Here, the important point is that the above matrices can be written in
terms of transformations in the Lorentz group.  In the two-by-two matrix
representation, the Lorentz boost along the $z$ direction takes the form
of Eq.(\ref{boost11}),
and the rotation along the $y$ axis can be written as Eq.(\ref{rot22}).
The boost along the $x$ axis takes the form
\begin{equation}
X(\chi) = \pmatrix{\cosh(\chi/2) & \sinh(\chi/2) \cr
\sinh(\chi/2) & \cosh(\chi/2)} .
\end{equation}
Then the core matrix of Eq.(\ref{core}) can be written as
\begin{equation}\label{phi11}
Z(\eta) R(\phi) Z(-\eta),
\end{equation}
or
\begin{equation}\label{phi22}
\pmatrix{\cos(\phi/2)  &
- e^{-\eta}\sin(\phi/2) \cr  e^{+\eta}\sin(\phi/2) & \cos(\phi/2) } ,
\end{equation}
if $ 1 < x < 2 $, where $Z(\eta)$ corresponds to a boost matrix along
the $z$ direction.

If $x$ is greater than 2, the upper-right element
of the core is positive and it can take the form
\begin{equation}\label{chi11}
Z(\eta) X(\chi) Z(-\eta),
\end{equation}
or
\begin{equation}\label{chi22}
 \pmatrix{\cosh(\chi/2)  &
 e^{-\eta}\sinh(\chi/2) \cr  e^{+\eta}\sinh(\chi/2) & \cosh(\chi/2) } .
\end{equation}

The expressions of Eq.(\ref{phi11}) and Eq.(\ref{chi11}) are a Lorentz
boosted rotation and a Lorentz-boosted boost matrix along the $x$
direction respectively.  These expressions play the key role in
understanding Wigner's little groups for relativistic particles~\cite{wig39}.

Wigner's little group is the maximal subgroup of the Lorentz group which
leaves the four-momentum of a given particle invariant.  If the particle is
massive, the little group is in the form of the three-dimensional rotation
group, as given in Eq.(\ref{phi11}) or Eq.(\ref{phi22}).  If the particle
has a space-like momentum, the little group is like Eq.(\ref{chi11}) or
Eq.(\ref{chi22}).

Let us look at their explicit matrix representations given
in Eq.(\ref{phi22}) and Eq.(\ref{chi22}).  The transition from
Eq.(\ref{phi22}) to Eq.(\ref{chi22}) requires the upper right element
going through zero.  This can only be achieved through $\eta$
going to infinity.  If we like to keep the lower-left element finite
during this process, the angle $\phi$ and the boost parameter $\chi$
have to approach zero.  The process of approaching the vanishing
upper-right element is necessarily a singular transformation.


\begin{table}
\caption{Massive and massless particles in one package.  Wigner's
little group unifies the internal space-time symmetries for massive and
massless particles.}\label{table11}
\vspace{3mm}
\begin{center}
\begin{tabular}{lccc}
\hline
{}&{}&{}&{}\\
{} & Massive, Slow \hspace{6mm} & COVARIANCE \hspace{6mm}&
Massless, Fast \\[4mm]\hline
{}&{}&{}&{}\\
Energy- & {}  & Einstein's & {} \\
Momentum & $E = p^{2}/2m$ & $ E = [p^{2} + m^{2}]^{1/2}$ & $E = p$
\\[4mm]\hline
{}&{}&{}&{}\\
Internal & $S_{3}$ & {}  &  $S_{3}$ \\[-1mm]
Space-time &{} & Wigner's  & {} \\ [-1mm]
Symmetry & $S_{1}, S_{2}$ & Little Group & Gauge Trans. \\[4mm]\hline
\end{tabular}
\end{center}
\end{table}

This limiting process, called group contraction~\cite{inonu53}, plays
the key role in unifying the internal
space-time symmetries of massive and massless particles.  This is like
Einstein's $E = \sqrt{(pc)^{2} + m^{2}c^{4}}$ becoming $E = pc$ in the
limit of large momentum or zero mass, as illustrated in Table~\ref{table11}.
This aspect of internal space-time symmetry has been discussed extensively
in the literature.  Table \ref{table11} represents a further content of
Einstein's energy-momentum relation~\cite{kiwi90jm}.

On the other hand, the core matrix of Eq.(\ref{core}) is an analytic
function of the variable $x$.  Thus, the lens matrix allows a
parametrization which allows the transition from massive particle to
massless particle analytically.  The lens optics indeed serves as the
analogue computer for this important transition in particle physics.

From the mathematical point of view,  Eq.(\ref{phi22}) and Eq.(\ref{chi22})
represent circular and hyperbolic geometries, respectively.  The
transition from one to the other is not a trivial mathematical procedure.
It requires a further investigation.

Let us go back to the core matrix of Eq.(\ref{core}).  The $x$ parameter
does not appear to be a parameter of Lorentz transformations.  However,
the matrix can be written in terms of another set of Lorentz transformations.
This aspect has been discussed in the literature.\cite{bk03}

\section*{Concluding Remarks}

The Lorentz group was introduced to physics by Einstein.  He was initially
interested in understanding Maxwell's equations.  The Lorentz group,
the language of Einstein, is the scientific language applicable to
all aspects of optical sciences, starting from Maxwell's equations.

\end{document}